\documentclass[twocolumn,showkeys,showpacs,preprintnumbers,prd,superscriptaddress,nofootinbib]{revtex4-1}
\usepackage{graphicx}
\usepackage{epsf}
\usepackage{bm}
\usepackage{amsmath}
\usepackage{amsfonts}
\usepackage{amssymb}
\usepackage{epstopdf}
\usepackage{color}
\usepackage{verbatim}
\usepackage{multirow}
\usepackage{bm}
\usepackage{soul}
\usepackage{hyperref}

\definecolor{darkpink}{rgb}{0.8, 0.0, 0.4}

\begin{document}

\title{
High-redshift transverse BAO measurements with the SDSS quasar catalog
}

\author{Felipe Avila}
\email{felipeavila@on.br}
\affiliation{Observatório Nacional, Rua General José Cristino 77, 
São Cristóvão, 20921-400 Rio de Janeiro, RJ, Brazil}

\author{Armando Bernui}
\email{bernui@on.br}
\affiliation{Observatório Nacional, Rua General José Cristino 77, 
São Cristóvão, 20921-400 Rio de Janeiro, RJ, Brazil}

\author{Miguel A. Sabogal}
\email{miguel.sabogalgarcia@unitn.it}
\affiliation{Department of Physics, University of Trento, Via Sommarive 14, 38123 Povo (TN), Italy}
\affiliation{Trento Institute for Fundamental Physics and Applications (TIFPA)-INFN, Via Sommarive 14, 38123 Povo (TN), Italy}

\author{Rafael C. Nunes}
\email{rafadcnunes@gmail.com}
\affiliation{Instituto de Física, Universidade Federal do Rio Grande do Sul, 91501-97 Porto Alegre, RS, Brazil}
\affiliation{Divisão de Astrofísica, Instituto Nacional de Pesquisas Espaciais, 
Avenida dos Astronautas 1758, 12227-010 São José dos Campos, SP, Brazil}

\begin{abstract}
Transverse Baryon Acoustic Oscillation (BAO) measurements offer a robust geometric probe of the Universe expansion, presenting minimal dependence on fiducial cosmological models. In this work, we analyze the SDSS-DR16 quasar catalog to search for the 2D BAO signal in the unexplored redshift interval 
$1.5 \leq z \leq 2.0$. 
Performing a fine tomographic analysis in 50 thin disjoint redshift shells ($\Delta z = 0.01$), to suppress line-of-sight projection smearing, and incorporating a full analytical covariance matrix, we detect the acoustic peak 
in two uncorrelated redshift shells: 
$\theta_{\rm BAO} = 1.911^{\circ} \pm 0.062^{\circ}$ and 
$\theta_{\rm BAO} = 1.727^{\circ} \pm 0.081^{\circ}$ 
centered at $z_{\rm eff} = 1.725$ and $z_{\rm eff} = 1.775$, with statistical significances of $3.4\,\sigma$ and $3.0\,\sigma$, respectively. 
By introducing a dimensionless shift parameter $\alpha$ in our empirical parameterization procedure, we then obtain two scaled angular diameter distances: $D_A/r_d = 11.00 \pm 0.36$ at $z_{\rm eff}=1.725$ and $D_A/r_d = 11.96 \pm 0.56$ at $z_{\rm eff}=1.775$. 
Incorporating these two novel data points into a literature compilation of 16 transverse BAO measurements, we perform a flat-$\Lambda$CDM parameter estimation, obtaining $\Omega_{m,0} = 0.41 \pm 0.06$ and $h r_d = 99.3 \pm 2.0$ Mpc.  
Our measurements successfully bridge a significant observational gap at high redshift, which remain highly consistent with the constraints reported by the Planck and DESI collaborations, demonstrating the potential of quasar tomographic surveys for dynamical dark energy studies.
\end{abstract}




\maketitle

\section{Introduction}
\label{sec:introduction}

Baryon Acoustic Oscillations (BAO) are a primordial phenomenon originating from sound waves in the early universe, which propagated until the decoupling of baryons and photons, leaving a characteristic geometrical imprint on the present-day distribution of cosmic structures~\citep{Peebles70, Einsenstein98,Meiksin1999}. 
The radius of these spherical sound waves reached the limit value $r_d$ at the end of the baryon drag epoch; this scale is known as the sound horizon. The spherical shell where the sound waves stopped due to decoupling left seeds of matter for the formation of structures. 
Therefore, tracers of matter (e.g., galaxies, quasars, HI) in the observed universe have an excess of probability to grow at the location of these spherical shells. 
The BAO scale, in this sense, can be used as a statistical standard ruler, an essential tool for measuring cosmic distances and, consequently, studying the nature of dark energy~\citep{Blake03,Seo03}. 

From an observational perspective, this excess of probability manifests in two-point statistics as oscillations in the power spectrum, $P(k)$~\citep{2dFGRS:2005yhx}, and as a peak in the two-point correlation function, $\xi(r)$~\citep{SDSS:2005xqv}.
In both three-dimensional estimators, detecting the BAO feature requires a large cosmological volume, a high number density of tracers, and precise spectroscopic redshifts~\citep{Blake03}. The BAO signal, whether measured in $\xi(r)$ or $P(k)$, arises from a combination of radial and transverse pairs of cosmic objects. Radial modes are sensitive to the Hubble parameter, $H(z)$, while transverse modes probe the angular diameter distance, $D_A(z)$. These components can be analyzed separately either along the line of sight~\citep{Gaztanaga09,Marra19} or across it~\citep{Sanchez11,Crocce2011,Carnero12,deCarvalho21}.
The transverse mode, in particular, is especially well suited for photometric surveys~\citep{Benitez:2008fs,Simpson09,Sanchez11}.

In an ideal situation, i.e., for an infinitesimally thin redshift shell, the transverse BAO measurement does not depend on any fiducial cosmological model, since the projection-effect correction in the BAO signal of the two-point angular correlation function (2PACF) is negligible.
In practice, however, finite surveys require the analysis to be performed using redshift shells of nonzero thickness, typically $\Delta z \sim 0.01$, and up to $\Delta z \sim 0.1$ in photometric catalogs.
As shown by \citet{Simpson09} and \citet{Sanchez11}, projection effects in this context shift the BAO peak toward smaller angular scales.
This effect can be mitigated in two ways: (i) by using thinner redshift shells, and (ii) by selecting high-redshift samples.
When applied together, these strategies yield an analysis that is effectively free of projection effects, eliminating the need for a shift correction based on a fiducial cosmology. Therefore, quasars represent ideal cosmic tracers for this type of model-independent analysis.

Quasars are important cosmological tracers~\citep{Song16}, but their low number density hinders clustering analyses due to high shot noise. 
Early studies in the 1980s, showed that quasars cluster strongly~\citep{Yee84,Shaver84,Shanks87,Bahcall91}, but robust statistical analyses only became possible with large-scale surveys like 2dF and SDSS~\citep{Croom:2004eg,SDSS03,Porciani04,Croom:2004ui,Myers07,daAngela:2006mf,Ross09}. 
These surveys confirmed that quasar clustering follows the linear model of structure formation, showing a time-evolving bias and a correlation function well described by a power law with $\gamma \simeq 1.5$ and $r_0 \simeq 5.0\,h^{-1}\mathrm{Mpc}$~\citep{Ross09, Zhao2021}. 
The increase in number density in current quasar catalogs ultimately enabled the first statistically significant detections of BAO in quasar samples about a decade later~\citep{Ata18}.



The transverse BAO signal has been detected in both the two-point~\citep{deCarvalho18} and three-point angular correlation functions~\citep{deCarvalho20}. 
Given the considerations discussed above regarding shell analysis, quasars present strong potential for probing cosmological parameters through transverse BAO measurements. 
However, determining the angular scale $\theta_{\rm BAO}$ remains challenging due to their low number density per square degree.
This limitation is critical, as the BAO feature in quasars is expected to appear at angular scales smaller than $2^{\circ}$.
Consequently, the entire survey area must be analyzed to achieve a statistically significant detection.
Using this strategy, \citet{deCarvalho18} measured $\theta_{\rm BAO} = 1.77^{\circ} \pm 0.31^{\circ}$, corresponding to a $2\sigma$ detection in the redshift range $2.20 \leq z \leq 2.25$, based on 10,526 quasars.
This result is consistent with the $\Lambda$CDM prediction within $1\sigma$, where the expected angular scale is $1.5^{\circ}$.

The primary objective of this work is the detection of a statistically significant transverse BAO signal in the redshift range of $1.5 \le z \le 2.0$, a regime where transverse measurements are currently lacking~\citep{Nunes20a,Menote22}. 
To achieve this goal, we perform a fine tomographic analysis utilizing very thin redshift shells of width $\Delta z = 0.01$. 
This narrow shell approach is crucial to minimize the smearing of the signal due to radial projection, 
thereby cleanly isolating the transverse BAO feature. 
Bridging this observational gap is of profound cosmological interest; a novel BAO anchor at that epoch, when combined with existing results, will enable more robust constraints on dark energy dynamics~\citep{Nunes20a, Bom2026}, provide new insights into the Hubble tension~\citep{Nunes20b}, and allow for stringent tests of possible dark sector interactions~\citep{Bernui23, Giare24}.

This work is organized as follows. In Section~\ref{sec:data_method}, we describe the data and methodology. The results are presented in Section~\ref{sec:Results}, and finally, in Section~\ref{sec:final}, we summarize our final remarks. 


\section{The Data set and Analysis methodology}
\label{sec:data_method}

High-precision cosmological analyses depend critically on large-scale structure catalogs that account for all potential systematic errors. Equally important is the construction of accurate random catalogs that mimics the survey's selection function, which are necessary to compute the 2PACF. 
The SDSS quasar catalog\footnote{Downloaded from: \url{https://data.sdss.org/sas/dr16/eboss/lss/catalogs/DR16/}} from the extended Baryon Oscillation Spectroscopic Survey Data Release 16 (eBOSS DR16) serves as a key example of such a resource~\citep{eBOSS:2020mzp}. It was produced specifically for cosmological analyses of BAO and Redshift-Space Distortions (RSD) using 2-point statistics.

The LSS quasar catalog provides a sample of 343,708 quasars within the redshift range $0.8<z<2.2$, selected for cosmological clustering analyses. The sample covers an effective sky area of $4,808$ deg$^{2}$. The catalog is divided into the North and South Galactic Cap (NGC and SGC). Both the data and random catalogs\footnote{The random catalog has fifty times more points than the quasar catalog in order to reduce noise~\citep{eBOSS:2020mzp}.} integrate these systematics and include statistical weights to optimize the cosmological measurements.

To ensure a robust analysis, the eBOSS DR16 quasar catalog employs a comprehensive weighting scheme. 
For our analyses, a series of weights is assigned to each quasar and random object, information designed to correct for specific observational biases~\citep{eBOSS:2020mzp}. 
This includes weights for fiber collisions ($w_{\rm cp}$), which address the inability to observe objects too close together on a single plate, and weights for redshift failures ($w_{\rm noz}$), which correct for spectrographic inefficiencies. Furthermore, systematic weights ($w_{\rm sys}$) are applied to mitigate correlations between the target density and imaging properties, such as stellar density and Galactic extinction. Finally, Feldman-Kaiser-Peacock (FKP) weights ($w_{\rm FKP}$)~\citep{Feldman94} are used to optimally balance the signal-to-noise ratio across the survey volume. As recommended, the total weight applied for clustering measurements is the product of these four corrections: $w_{\rm tot} = w_{\rm sys} \cdot w_{\rm cp} \cdot w_{\rm noz} \cdot w_{\rm FKP}$.

For our analysis using thin redshift shells, the variation of these weights within a single shell is negligible. 
Therefore, while we apply the recommended weights, their effect on our measurements is minimal.

For our detailed search for the transverse BAO signature, we perform a tomographic scan using 50 thin disjoint redshift bins of width 
$\Delta z = 0.01$. 
We focused our blind search primarily on the intermediate interval 
$1.50 \le z \le 2.00$, where the quasar number density is optimally high for clustering measurements, yet sufficiently separated from the existing high-redshift measurement at $z=2.225$~\citep{deCarvalho18} to ensure an independent transverse BAO measurement for constraints on cosmological parameters. 
As discussed in Section~\ref{sec:Results}, this rigorous scan robustly identifies the BAO signatures 
in four thin redshift shells in the interval $z \in [1.7, 1.8]$.

\subsection{Landy-Szalay estimator}
\label{sec:LS_estimator}

The two-point angular correlation function (2PACF) is measured using the Landy-Szalay (LS) estimator~\citep{Hewett82,Landy93}. 
This estimator has advantageous statistical properties: it is unbiased, exhibits minimal variance, and returns the smallest deviations for a given cumulative probability compared to other estimators~\citep{Kerscher00}. The LS estimator for the 2PACF, denoted $\omega(\theta)$, is defined by the expression~\citep{Landy93} 
\begin{equation}
\omega(\theta) \equiv \frac{DD(\theta) - 2DR(\theta) + RR(\theta)}{RR(\theta)}\,,
\end{equation}
where $DD(\theta)$, $RR(\theta)$, and $DR(\theta)$ are the normalized counts of data-data, random-random, and data-random pairs separated by an angle $\theta$.

The angular separation $\theta$ between any two objects on the celestial sphere is calculated from their right ascension ($\alpha$) and declination ($\delta$) coordinates using the great-circle distance formula 
\begin{equation}
\theta = \arccos[\sin \delta_A \sin \delta_B + \cos \delta_A \cos \delta_B \cos(\alpha_A-\alpha_B)] \,.
\end{equation}
The pair counts and the computation of $\omega(\theta)$ are efficiently calculated using the \texttt{TreeCorr}\footnote{\url{https://rmjarvis.github.io/TreeCorr/_build/html/index.html}} code~\citep{Jarvis04}, which employs a tree-based algorithm to handle large datasets.

\subsection{Covariance matrix of the 2PACF}\label{sec:ln_sim}

To evaluate the statistical significance of the BAO signal, we employ an analytical model for the covariance matrix of the 2PACF. 
Sample covariances can be estimated from a large suite of mock catalogs. 
However, the production of high-quality quasar mocks that accurately reproduce the survey mask and non-linear clustering at high redshifts is a difficult task~\citep{Smith20}. 
Fortunately, for the very thin tomographic shells, with $\Delta z = 0.01$ as considered in our analysis, the clustering signal operates predominantly in the linear regime and the error budget has a major contribution from shot-noise. In this limit, the analytical Gaussian covariance, which self-consistently couples the Poisson noise with the fiducial clustering $C_\ell$ terms, provides a highly accurate and computationally efficient description of the errors, in excellent agreement with $N$-body simulations~\citep{Crocce2011, Sanchez11, Abbott24}.

Accounting for partial sky coverage, $f_{\mathrm{sky}}$, and the shot-noise term, $1/\bar{n}$, the theoretical Gaussian covariance matrix is defined as~\citep{Crocce2011} 
\begin{equation}\label{eq:cov_matrix}
\begin{split}
\text{Cov}(\theta, \theta') \equiv \, & \frac{2}{f_{\mathrm{sky}}} \sum_{\ell \ge 0} \frac{2\ell+1}{(4\pi)^2} P_\ell(\cos\theta) P_\ell(\cos\theta') \\
& \times \left( b^{2}_{\rm Q} \,C_\ell + \frac{1}{\bar{n}} \right)^{2} \,,
\end{split}
\end{equation}
where $P_\ell(\cos\theta)$ are the Legendre polynomials evaluated at the centers of the respective angular bins, and $\{ C_\ell \}$ is the angular power spectrum from the fiducial cosmological model. The power spectrum is computed using the public code \texttt{CCL}\footnote{\url{https://github.com/LSSTDESC/CCL}}~\citep{Chisari19}, assuming the baseline \textit{Planck} 2018 cosmology (TT, TE, EE+lowE)~\citep{Planck20}.

Quasars are highly biased tracers of the large-scale structure. To compute the $b^{2}_{\rm Q} C_\ell$ term, we adopt the redshift-dependent linear bias parameterization proposed by \citet{Laurent17} 
\begin{equation}
b_{\rm Q}(z) = \hat{\alpha} \left[(1+z)^2 - 6.565\right] + \hat{\beta},
\end{equation}
with the best-fit parameters $\hat{\alpha} = 0.278$ and $\hat{\beta} = 2.393$. 
This prescription is consistent with recent high-redshift structure formation analyses~\citep{Piccirilli:2024xgo}. Instead of adopting a global bias, we evaluate $b_{\rm Q}(z)$ at the effective mid-point redshift of each individual tomographic shell to construct their specific covariance matrices prior to the parameter estimation.

\subsection{Measuring the angular BAO signal}
\label{eq:Sanchez_method}

In the analysis of the transverse BAO signal, the excess probability of finding correlated pairs of quasars is revealed by the 2PACF. The measured scale of this correlation, which we denote as $\theta_{\rm FIT}$, is not exactly the theoretical transverse BAO scale, $\theta^{\rm th}_{\rm BAO}$. This is because the non-zero thickness of the redshift bin containing the data can produce a shift in the acoustic signature due to the projection effect~\citep{Sanchez11,deCarvalho18}. However, if our data are contained in a very thin shell at high redshift, the projection effect is negligible (see Appendix~\ref{app:proj_effect}). In such a case, one expects $\theta_{\rm FIT} \simeq \theta^{\rm th}_{\rm BAO}$, where the percentage difference between these quantities is less than a fraction of $1\%$~\citep{Sanchez11}.

The standard approach to isolating the BAO feature is to fit the correlation function with a parameterization that combines a smooth broadband model with a Gaussian peak~\citep{Smith:2007gi,Sanchez11}: the former accounts for the overall shape at different scales, while the latter describes the acoustic peak and its width. Standard power laws often fail to capture the background when the 2PACF exhibits negative correlations around the BAO scale. To address this, we introduce a more flexible parameterization using a second-degree polynomial for the continuum.

Furthermore, to directly quantify the deviation of the measured acoustic scale from the fiducial cosmology, we introduce the dimensionless shift parameter $\alpha$, defined such that $\theta_{\rm FIT} = \alpha\,\theta^{\rm th}_{\rm BAO}$. Our full parameterization of $\omega(\theta)$ is therefore 
\begin{equation}\label{eq:ajuste}
\omega(\theta) = a + b\,\theta + c\,\theta^{2} + C\exp\left[-\frac{(\theta-\alpha\,\theta^{\rm th}_{\rm BAO})^2}{2\,\sigma^{2}}\right] \,,
\end{equation}
where $a$, $b$, $c$, $C$, $\alpha$, and $\sigma$ are free parameters.\footnote{A similar methodology can be applied for a 3D analysis. Using a 5th order polynomial fit, it is possible to obtain the linear point, i.e., a scale with a correlation function that depends weakly on the non-linear evolution of the structures~\citep{Anselmi22,Lee:2024rvh}.} We adopt the following broad flat priors: $a \in [-1,1]$, $b \in [-1,1]$, $c \in [-1,1]$, $\alpha \in [0.5, 1.5]$, $C \in [0,0.5]$ and $\sigma \in [0.0^\circ, 0.5^\circ]$.

In our analyses, the main observable is the transverse BAO angular scale, given by $\theta_{\rm FIT}(z)$. 
Once extracted, this angular scale can be translated into a constraint on the angular diameter distance, $D_A(z)$, by means of the purely geometric relationship 
\begin{equation}\label{eq:da_bao}
\frac{D_{A}(z)}{r_{d}} = \frac{1}{(1+z)\, \theta_{\rm FIT}(z)} \,,
\end{equation}
where $r_d$ is the sound horizon at the drag epoch and $\theta_{\rm FIT}$ is expressed in radians. This relation allows us to derive cosmological distance constraints at high redshift with minimal assumptions about the underlying fiducial cosmology, as we shall explore in the subsequent section.

\section{Main Results}
\label{sec:Results}

Thin shell analyses of the transverse BAO signature aim to robustly isolate the acoustic feature while minimizing line-of-sight projection smearing. However, unlike full 3D analyses or wide-bin projections, the number of quasars per square degree drastically decreases when a very narrow redshift bin, i.e. $\Delta z = 0.01$, is considered. 
Consequently, the BAO peak in the angular correlation function is generally measured with a low signal-to-noise ratio. 
In practice, this low number density makes the search for the transverse BAO signal in quasar catalogs particularly challenging. 

Considering that there is already the transverse BAO measurement at $z=2.225$~\citep{deCarvalho18}, we targeted the unexplored redshift interval $1.50 \le z \le 2.00$. 
By performing a full MCMC analysis across all 50 thin redshift shells in this interval, we identified four bins that yield a local statistical significance of at least $3\,\sigma$. Notably, these high-significance detections emerge as two pairs of contiguous redshift bins: $z = 1.715,\, 1.725$ and $z = 1.775,\, 1.785$.

\subsection{BAO fit and statistical significance}
\label{sec:bao_fit}

To constrain the free parameters of equation~(\ref{eq:ajuste}) from the correlation functions of our four selected tomographic shells, we employed a Markov Chain Monte Carlo (MCMC) approach using the \texttt{emcee} sampler~\citep{Foreman-Mackey_2013}, adopting the uniform priors previously defined in Section~\ref{eq:Sanchez_method}. This method efficiently explores the multidimensional parameter space and provides robust estimates of the marginalized posterior distributions. The fits are performed over the full angular interval $\theta \in [0.0^\circ, 3.0^\circ]$.

The theoretical covariance matrices were computed using equation~(\ref{eq:cov_matrix}) with an effective quasar number density of $\bar{n} = 1826.3~{\rm sr}^{-1}$ and a partial sky coverage of $f_{\rm sky}=0.12$. The linear bias for each shell was evaluated at its effective mid-point redshift, yielding $b_{\rm Q}(1.715) = 2.617$, $b_{\rm Q}(1.725) = 2.632$, $b_{\rm Q}(1.775) = 2.709$, and $b_{\rm Q}(1.785) = 2.724$. 
We note that while the Poisson noise is significant for such thin slices, the full theoretical covariance yields errors that are noticeably larger than those derived strictly from pure shot-noise estimates. 
Clearly, relying solely on the shot-noise limit would underestimate the uncertainties and artificially inflate the statistical significance of the detections. 
Hence, the inclusion of the cosmological clustering terms in the analytical covariance is essential to avoid biasing the results 
(see section~\ref{sec:ln_sim}).

\begin{figure*}[htbp]
\centering
\includegraphics[width=0.48\textwidth]{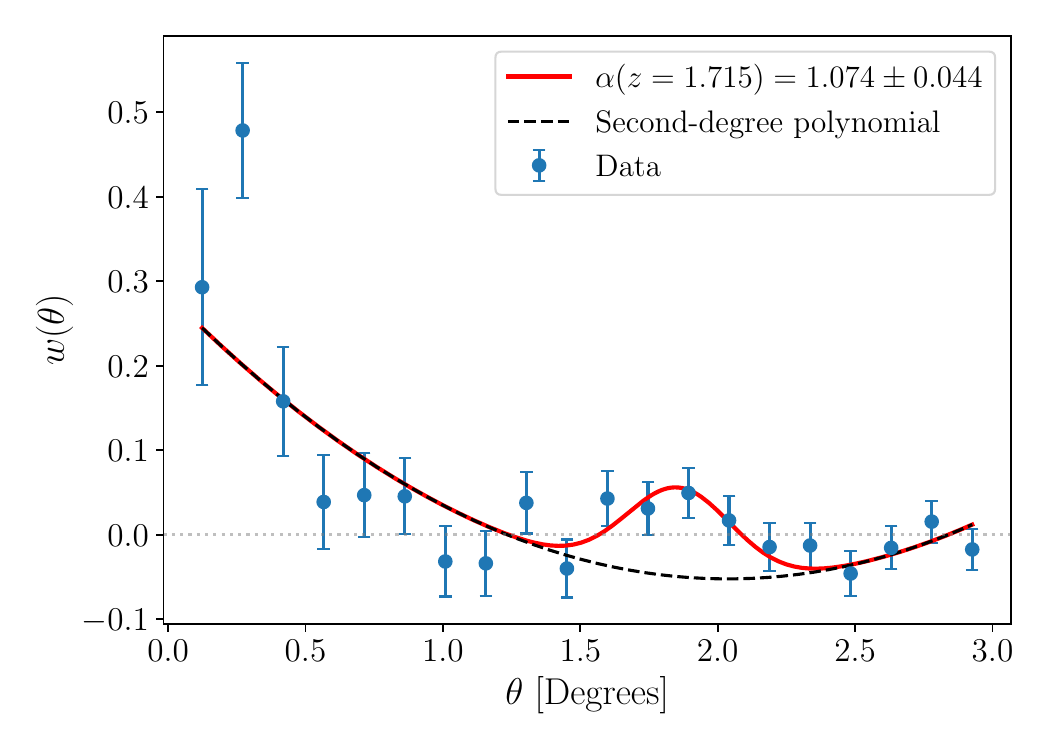}\hfill
\includegraphics[width=0.48\textwidth]{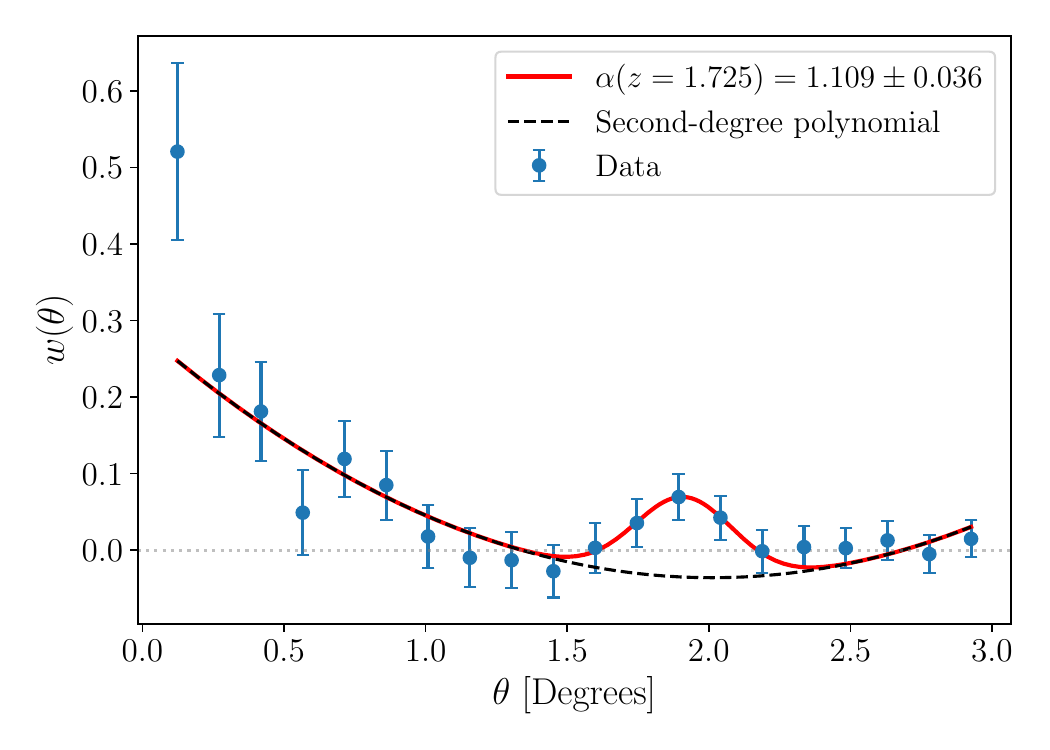}
\vspace{0.4cm} 
\includegraphics[width=0.48\textwidth]{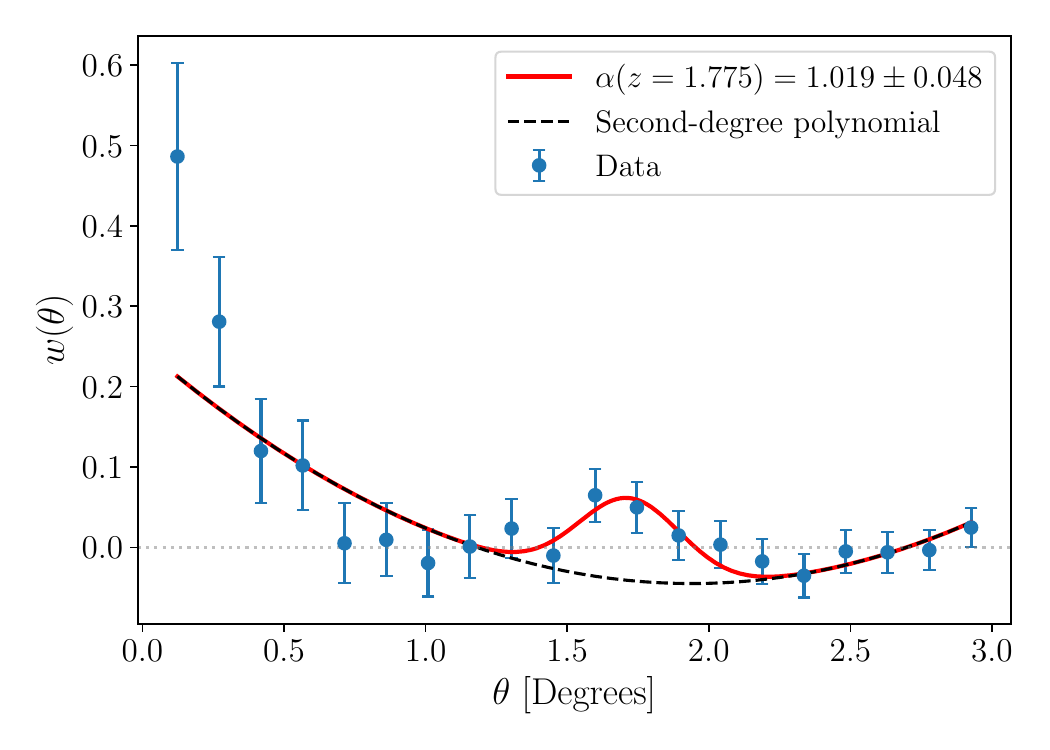}\hfill
\includegraphics[width=0.48\textwidth]{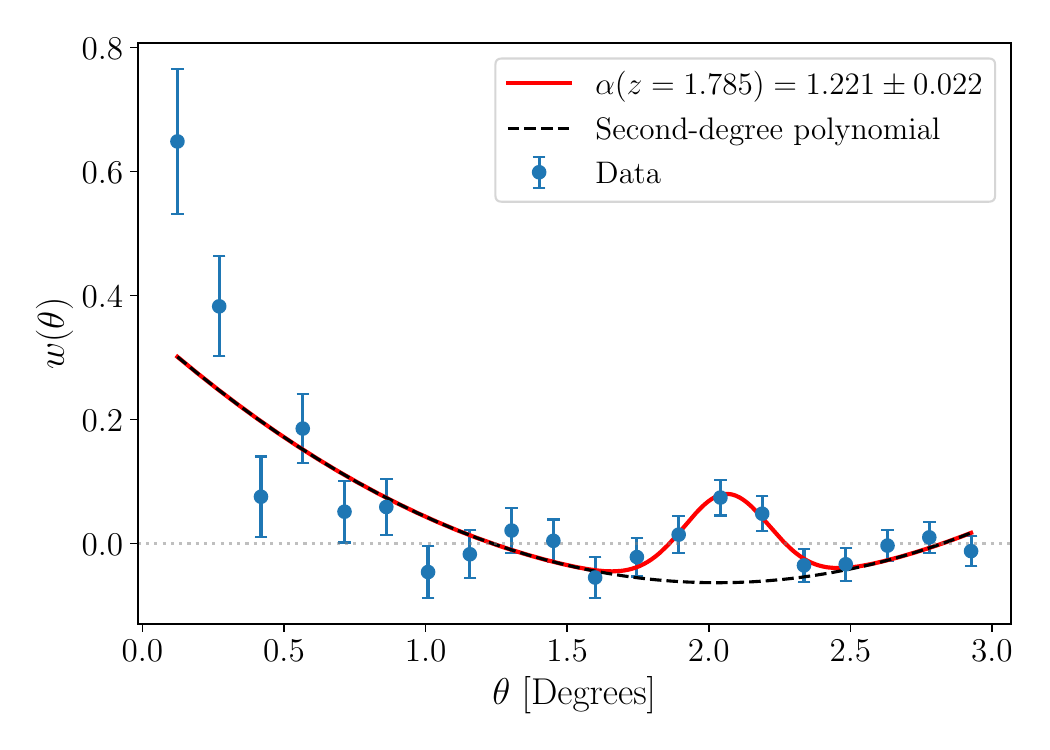}
\caption{Best-fit models for the 2PACF, $w(\theta)$, of the SDSS quasar sample at four selected tomographic bins: $z_{\rm eff} = 1.715$ and $1.725$ (top row), and $z_{\rm eff} = 1.775$ and $1.785$ (bottom row). The error bars are derived from the square root of the diagonal elements of the full theoretical covariance matrix. In each panel, the red solid line represents the full BAO best-fit, while the dashed black line indicates the smooth polynomial background without BAO signal. 
The reduced-$\chi^2$ values, $\chi_{\nu}^2$, 
for $z_{\rm eff} = 1.715, 1.725, 1.775, 1.785$ are $\chi_{\nu}^2 = 2.13, 0.95, 1.04, 2.10$, respectively. 
} 
\label{fig:all_fits}
\end{figure*}

The empirical detection of the acoustic peak is first quantified by comparing the minimum $\chi^2$ 
of the full BAO model, $\chi^2_{\rm BAO}$, 
to the minimum $\chi^2$ of the null model, $\chi^2_{\rm null}$, i.e., obtained by forcing $C=0$ 
in equation~(\ref{eq:ajuste}). 
The statistical significance of the BAO detection in the 2PACF, in units of Gaussian standard 
deviations ($\sigma$), is obtained from $\Delta\chi^2 = \chi^2_{\rm null} - \chi^2_{\rm BAO}$. 
The MCMC pipeline yields the following constraints for the shift parameter $\alpha$ and their 
respective local significances: $\alpha = 1.074 \pm 0.044$ ($3.18\,\sigma$) at $z_{\rm eff} = 1.715$; $\alpha = 1.109 \pm 0.036$ ($3.40\,\sigma$) at $z_{\rm eff} = 1.725$; $\alpha = 1.019 \pm 0.048$ ($3.00\,\sigma$) at $z_{\rm eff} = 1.775$; and $\alpha = 1.221 \pm 0.022$ ($4.79\,\sigma$) at $z_{\rm eff} = 1.785$. 
The best-fit results, overlaid with the observed data points for these four shells, are displayed in Figure~\ref{fig:all_fits}. 
The corresponding full marginalized posterior contour plots are reserved for Appendix~\ref{app:contours}.

We would like to point out that these fits reveal a highly consistent extraction of the BAO feature across multiple adjacent bins. The sub-samples at $z_{\rm eff} = 1.715$, $1.725$, and $1.775$ exhibit robust physical parameters, with their $\alpha$ constraints closely tracking the fiducial $\Lambda$CDM prediction ($\alpha \simeq 1$). 
However, although the shell at $z_{\rm eff} = 1.785$ formally exhibits the highest mathematical significance, i.e. $4.79\,\sigma$, its best-fit shift parameter ($\alpha = 1.221 \pm 0.022$) is anomalously high. 
Our criterion for ultimately deciding on a robust transverse BAO measurement involves calculating the goodness-of-fit for the 2PACF data, selecting only those redshift shells with exceptional reduced chi-squared values, i.e., $\chi^2_\nu \simeq 1.0$. 
This will be investigated in the next subsection.



\subsection{Cosmological interpretation}
\label{sec:cosmology}

In this section, we perform statistical analyses of a public dataset of transverse BAO measurements, with the inclusion of two new measurements obtained in this work, in order to determine the best-fit parameters $(\Omega_m, h r_d)$ of the flat $\Lambda$CDM model.

As discussed in Section~\ref{sec:bao_fit}, our tomographic analysis identified four shells with highly significant BAO signature. 
However, adjacent bins (e.g., $z_{\rm eff} = 1.715$ and $1.725$) exhibit intrinsic statistical correlation due to their proximity. 
We select for our analyses non-overlapping redshift bins, to obtain uncorrelated BAO measurements. 
Furthermore, evaluating the goodness-of-fit for our 6-parameter model (see equation~(\ref{eq:ajuste})) over 20 angular data points (i.e., with 14 degrees of freedom), we find that the shells at $z_{\rm eff} = 1.725$ and $z_{\rm eff} = 1.775$ provide exceptional reduced chi-squared values ($\chi^2_\nu \simeq 0.95$ and $\chi^2_\nu \simeq 1.04$, respectively), whereas the bins at $z_{\rm eff}= 1.715$ and $z_{\rm eff} = 1.785$ present noisier fits ($\chi^2_\nu > 2.0$). 
Since the two selected optimal shells --at $z_{\rm eff} = 1.725$ and $z_{\rm eff} = 1.775$-- 
have a redshift separation of $0.05$ (a value much larger than the thickness of our shells, 
$\Delta z = 0.01$), they do not overlap and can therefore be safely considered independent measurements.

Our joint analysis therefore considers a total of 18 independent $\{ \theta_{\rm BAO}(z) \}$ measurements: 14 data points from~\cite{Menote22} in the redshift range $0.35 \leq z \leq 0.63$, one measurement from~\cite{deCarvalho18} at $z_{\rm eff} = 2.225$, one from~\cite{deCarvalho21} at $z_{\rm eff} = 0.11$, and the two robust measurements obtained in this work at $z_{\rm eff} = 1.725$ and $z_{\rm eff} = 1.775$.

\begin{figure}[htbp]
\centering
\includegraphics[width=0.48\textwidth]{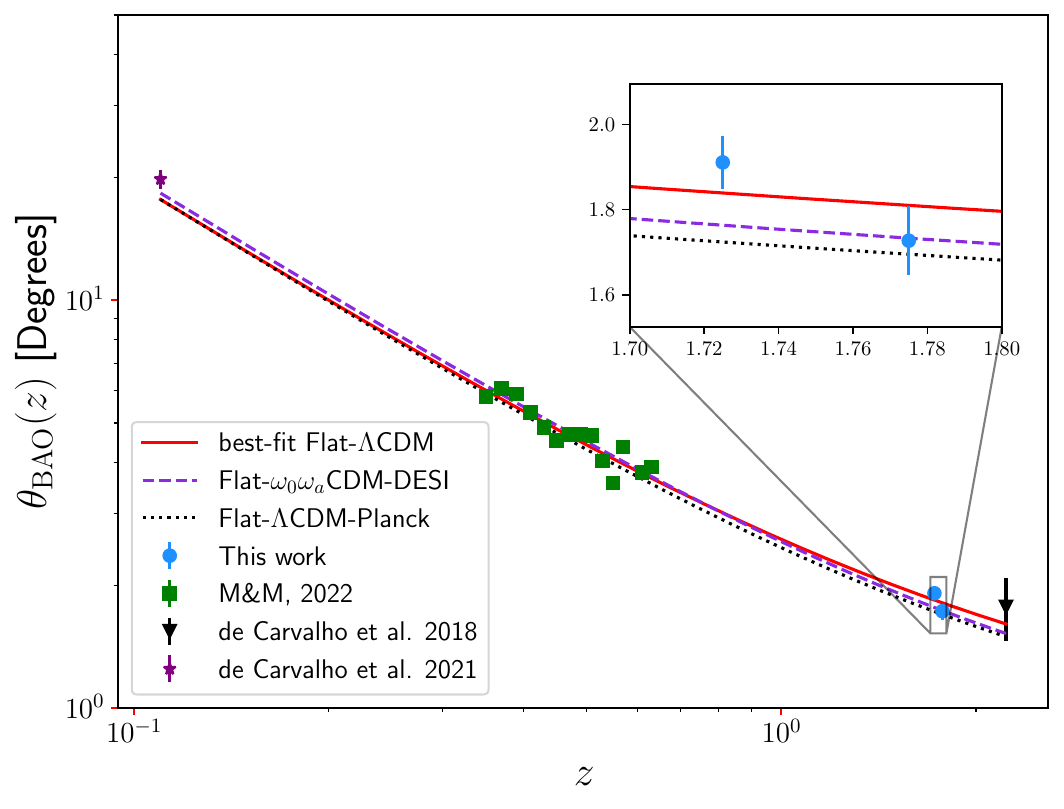}
\caption{The theoretical predictions for the $\theta_{\rm BAO}(z)$ function derived for the $\Lambda$CDM and $w_0w_a$CDM models, using the fixed best-fit parameter values $(\Omega_{m,0}, h r_d)$ reported in Table~\ref{table-results}. 
The curves are compared with the 18 $\theta_{\rm BAO}$ measurements available in the literature, 
including two new uncorrelated quasar measurements at $z_{\rm eff} = 1.725$ and $z_{\rm eff} = 1.775$ obtained in this work (the zoomed-in area shows these data).
}
\label{fig:bestfit-bao}
\end{figure}

Table~\ref{table-results} summarizes the best-fit parameters obtained by fitting the theoretical 
$\theta_{\rm BAO}(z)$ function to the dataset of 18 transverse BAO measurements. 
The corresponding best-fit curve overlaid with the observational points is illustrated in Figure~\ref{fig:bestfit-bao}. 
As shown in this table, our constraints on the matter density ($\Omega_{m,0} = 0.407 \pm 0.063$) and the scaled sound horizon ($h r_d = 99.25 \pm 2.04$ Mpc) are fully consistent with previous analyses in the literature, maintaining good agreement with independent surveys such as \textit{Planck}~\citep{Planck20} and DESI~\citep{DESI25}. 
This agreement reinforces the robustness of transverse BAO measurements as a highly competitive dataset for probing the expansion history of the Universe. 

\begin{table}[htbp]
\centering
\renewcommand{\arraystretch}{1.3} 
\begin{tabular}{|l|c|c|}
\hline
Model $\backslash$ Parameters & $\Omega_{m,0}$ & $h r_d$ [Mpc] \\
\hline
$\Lambda$CDM (\textit{Planck}) & $0.3166 \pm 0.0084$ & $98.92 \pm 0.91$ \\
$w_0 w_a$CDM (DESI) & $0.2975 \pm 0.0086$ & $101.54 \pm 0.73$ \\
$\Lambda$CDM (\textbf{this work}) & $0.407 \pm 0.063$ & $99.25 \pm 2.04$ \\
\hline
\end{tabular}
\caption{
Cosmological parameters found by the \textit{Planck}~\citep{Planck20} and DESI~\citep{DESI25} collaborations in their corresponding analyses. 
For comparison, we display the best-fit values of the $\Lambda$CDM parameters 
$(\Omega_{m,0}, h r_d)$ that best fit the set of 18 transverse $\theta_{\rm BAO}$ measurements, incorporating the two independent high-redshift data points obtained in this work. 
The result of this best-fitting procedure is shown in Figure~\ref{fig:bestfit-bao}.} 
\label{table-results}
\end{table}

\section{Final Remarks}
\label{sec:final}

Baryon Acoustic Oscillations have established themselves as one of the most robust cosmological probes for investigating the accelerated expansion of the Universe. Recent measurements from the DESI collaboration provide compelling evidence in favor of dynamical dark energy~\citep{DESI25}. Nevertheless, the Hubble tension remains unresolved, highlighting the need for new and precise observations to clarify these outstanding issues~\citep{CosmoVerseNetwork:2025alb}.

Measurements of the transverse signal of the BAO phenomenon at several redshifts, 
offer a valuable methodology for testing cosmological models with minimal assumptions about the fiducial background~\citep{Sanchez11, Menote22, Nunes20a}. 
In this work, we analyzed the SDSS quasar catalog searching for transverse BAO signals in thin high-redshift shells. 
Indeed, at high redshifts and for shells with $\Delta z = 0.01$, the line-of-sight projection effect is strictly negligible~\citep{Sanchez11}, allowing us to obtain transverse BAO measurements that are statistically significant and independent of the radial physics (see Appendix~\ref{app:proj_effect}).

In fact, obtaining a new transverse BAO anchor at high redshift is particularly important because there was a significant gap in the literature between $z=0.63$~\citep{Menote22} and $z=2.225$~\citep{deCarvalho18}. 
In this study, we successfully bridged this gap by identifying two robust transverse BAO signals in the narrow redshift shells centered at $z_{\rm eff} = 1.725$ and $z_{\rm eff} = 1.775$.

By employing a full analytical covariance matrix to account for both shot-noise and clustering variance, and by utilizing the dimensionless shift parameter $\alpha$, we extracted the acoustic scale with local statistical significances of $3.4 \sigma$ and $3.0 \sigma$ for the measurements at $z_{\rm eff} = 1.725,\, 1.775$, respectively. 
The corresponding angular diameter distances for these statistically significant 
--and uncorrelated-- bins are displayed in Table~\ref{table-final-results}. 

\begin{table}[htbp]
\centering
\renewcommand{\arraystretch}{1.3} 
\begin{tabular}{|l|c|c|c|}
\hline
redshift & $\theta_{\rm BAO}$ & $D_A / r_d$ & Stat. Significance \\
\hline
$z_{\rm eff}$ = $1.725$ & $1.911^\circ \pm 0.062^\circ$ & $11.003 \pm 0.357$ & $3.4\,\sigma$ \\
$z_{\rm eff}$ = $1.775$ & $1.727^{\circ} \pm 0.081^{\circ}$ & $11.959 \pm 0.563$ & $3.0\,\sigma$ \\
\hline
\end{tabular}
\caption{
Summary of our results: two statistically significant transverse BAO measurements, at 
high redshifts, using the SDSS quasar catalog.} 
\label{table-final-results}
\end{table}

These new high-redshift constraints provide valuable transverse BAO data points in the redshift interval $1.5 \le z \le 2.0$, offering improved leverage for mapping the expansion history of the Universe. 

To investigate the consistency of our measurements, in Section~\ref{sec:cosmology} we incorporate the two transverse BAO data points into a public compilation, expanding 
the list of these BAO data to a total of 18 transverse BAO measurements. 
A statistical MCMC analysis within the flat-$\Lambda$CDM framework was then performed to determine the best-fit parameters $(\Omega_{m,0}, h r_d)$. As shown in Figure~\ref{fig:bestfit-bao}, our joint fit yields $\Omega_{m,0} = 0.407 \pm 0.063$ and $h r_d = 99.25 \pm 2.04$ Mpc.

Our constraints are in excellent agreement with similar analyses that study the $\Lambda$CDM 
and the $w_0w_a$CDM models. 
This comparison confirms the compatibility of our high-redshift transverse BAO detections with existing cosmological constraints, proving that quasar tomographic surveys can competitively constrain the dark sector.

It is equally important to comment on the ongoing debate regarding whether transverse $\theta_{\rm BAO}$ measurements are in tension with the predictions of the standard $\Lambda$CDM model~\citep{Bernui23,Dwivedi24,Favale24,Zheng25,Favale2026,Pantos2026,Kumar2026}. 
However, depending on the dataset considered, no significant tension is observed~\citep{Menote22}, a conclusion supported by our updated analysis. A more conclusive assessment of this question requires a full Bayesian hierarchical analysis that consistently accounts for uncertainties and cross-correlations across disparate datasets. Such a comprehensive study is currently being presented in a companion work~\citep{Miguel_ON}.



Finally, for the sake of completeness and transparency, we note that a preliminary version of this analysis (made available as a preprint) reported an earlier measurement for the $z_{\rm eff}=1.725$ shell, yielding $\theta_{\rm BAO} = 1.928^\circ \pm 0.094^\circ$. 
This preliminary measurement was adopted in the recent companion study by \citet{Miguel_ON}. 
Thanks to the refined methodology presented in this final version --specifically the implementation of the full analytical covariance matrix and the dimensionless shift parameter $\alpha$-- our updated and definitive measurement for this bin corresponds to $\theta_{\rm BAO} = 1.911^\circ \pm 0.062^\circ$. As both values are statistically equivalent (well within $1\sigma$), the cosmological implications and the broader Bayesian analysis discussed in \citet{Miguel_ON} remain fully robust and unaffected by the updated results.

Future progress in transverse BAO measurements will greatly benefit from ongoing and upcoming large-scale structure surveys. 
Projects such as DESI~\citep{DESIDR1}, Euclid~\citep{Euclid}, and LSST~\citep{LSST} will provide unprecedented spectroscopic and photometric datasets, enabling precise determinations of the BAO scale over a wide redshift range. 
These new and precise data will not only improve the current available compilation, but also allow for more stringent tests of alternative cosmological scenarios, including dynamical dark energy.

\begin{acknowledgments}

FA acknowledges Fundação Carlos Chagas Filho de Amparo à Pesquisa do Estado 
do Rio de Janeiro (FAPERJ), Processo SEI-260003/001221/2025, for the financial 
support. AB acknowledges a CNPq fellowship. 
M.A.S acknowledges support from CAPES and expresses gratitude to the Observatório Nacional for their hospitality during the development of this work. 
M.A.S also acknowledges support from the Istituto Nazionale di Fisica Nucleare (INFN) through the Commissione Scientifica Nazionale 4 (CSN4) Iniziativa Specifica ``Quantum Fields in Gravity, Cosmology and Black Holes'' (FLAG). 
R.C.N. thanks the financial support from the Conselho Nacional de Desenvolvimento Científico e Tecnológico (CNPq, National Council for Scientific and Technological Development) under the project No. 304306/2022-3, and the Fundação de Amparo à Pesquisa do Estado do RS (FAPERGS, Research Support Foundation of the State of RS) for partial financial support under the project No. 23/2551-0000848-3. 



\end{acknowledgments}

\section*{Data Availability}
The data used in this work are available from the corresponding author upon reasonable request.

\bibliographystyle{apsrev4-1}
\bibliography{main}

\appendix

\section{Projection effects}
\label{app:proj_effect}

As established in \citet{Sanchez11}, projection effects are significantly suppressed at high redshifts. To verify that this behavior holds for our tomographic selection, we performed a theoretical check by computing the theoretical 2PACF for different shell thicknesses centered at $z_{\rm eff} = 1.725$. Given the narrow range of our tomographic scan ($z \in [1.5, 2.0]$) and the identical shell widths used, this calculation serves as a representative baseline for all our analyzed bins.

The theoretical angular correlation is computed as:
\begin{equation}
\omega(\theta) = \int dz_1 \phi(z_1) \int dz_2 \phi(z_2) \xi(r(z_1), r(z_2), \theta) \,,
\end{equation}
where $\phi(z)$ is the normalized radial selection function and $\xi(r)$ is the spatial correlation function, computed using the \texttt{CCL} code~\citep{Chisari19}. 

\begin{figure}
\centering
\includegraphics[scale=0.6]{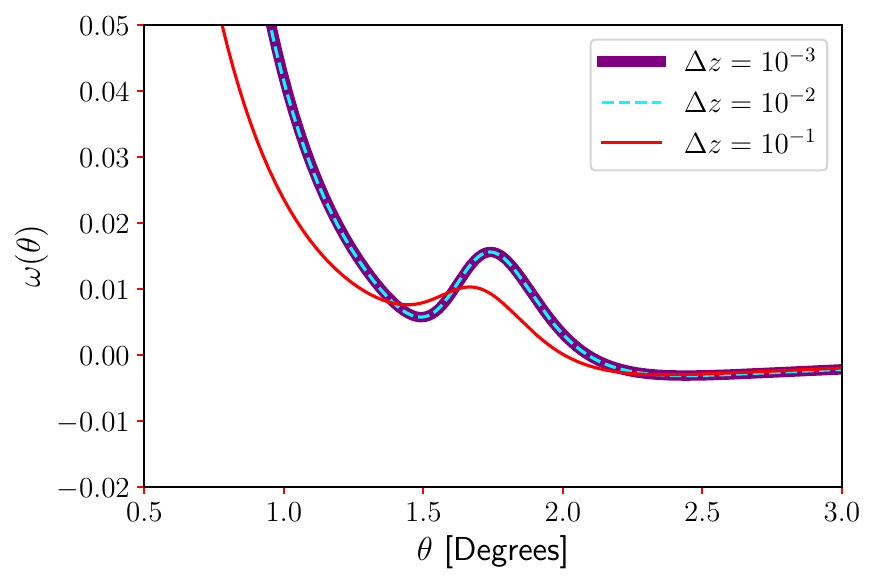}
\caption{Theoretical 2PACF $\omega(\theta)$ at $z = 1.725$ for redshift bin sizes: $\Delta z = 10^{-3}$, $10^{-2}$, and $10^{-1}$. The curves for $\Delta z = 10^{-3}$ and $\Delta z = 10^{-2}$ overlap, indicating that projection smearing is negligible for our tomographic analysis.}
\label{fig:theo_curve}
\end{figure}

Figure~\ref{fig:theo_curve} shows the theoretical $\omega(\theta)$ for three bin widths: $\Delta z = 0.001$, $0.01$, and $0.1$. The results demonstrate that the functions for $\Delta z = 0.001$ and $\Delta z = 0.01$ are essentially indistinguishable. This confirms that for our chosen shell width of $\Delta z = 0.01$, projection effects are negligible and require no further theoretical correction across the entire redshift range considered in this work. Significant deviations only emerge at $\Delta z = 0.1$, which further corroborates the findings of \citet{Sanchez11} regarding the robustness of thin-shell BAO analyses.

\section{Posterior contour plots}
\label{app:contours}

\begin{figure*}[htbp]
\centering
\includegraphics[width=0.85\textwidth]{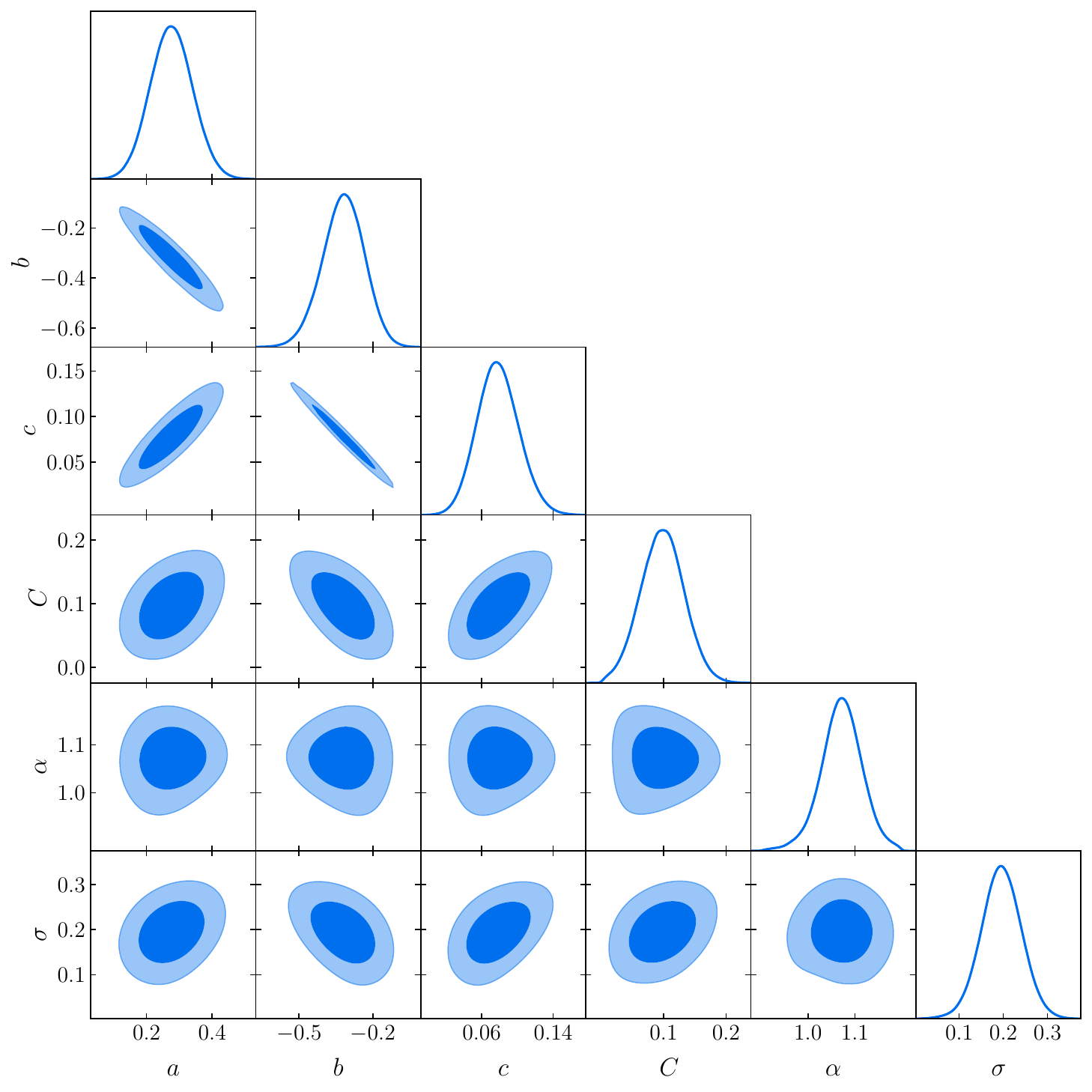}
\caption{Marginalized posterior distributions for the parameters of the transverse BAO model obtained through the MCMC analysis of the SDSS quasar sample for the tomographic bin at $z_{\rm eff} = 1.715$. The inner and outer contours represent the $68\%$ and $95\%$ confidence levels, respectively.}
\label{fig:contour_1715}
\end{figure*}

\begin{figure*}[htbp]
\centering
\includegraphics[width=0.85\textwidth]{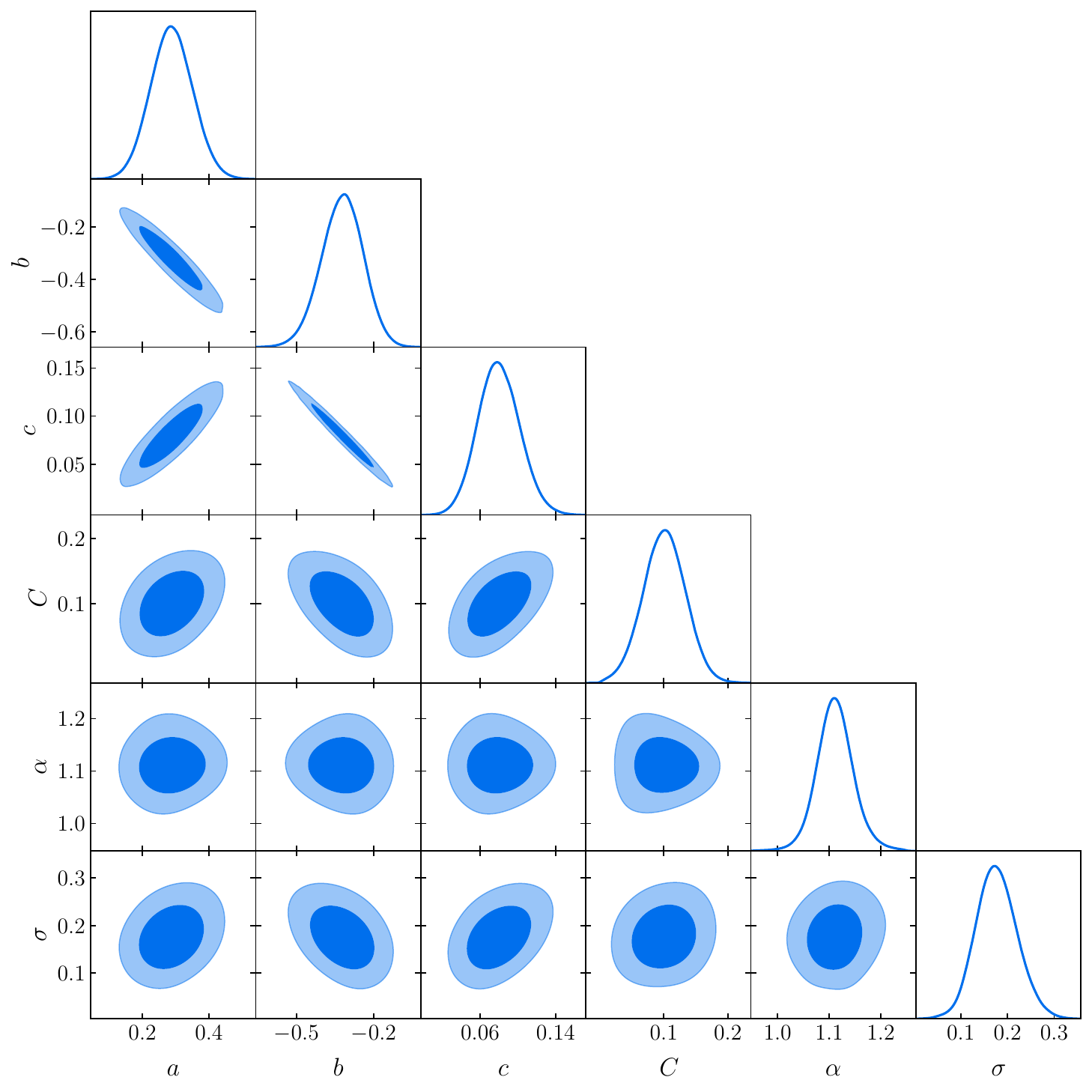}
\caption{Same as Figure~\ref{fig:contour_1715}, but for the tomographic bin at $z_{\rm eff} = 1.725$.}
\label{fig:contour_1725}
\end{figure*}

\begin{figure*}[htbp]
\centering
\includegraphics[width=0.85\textwidth]{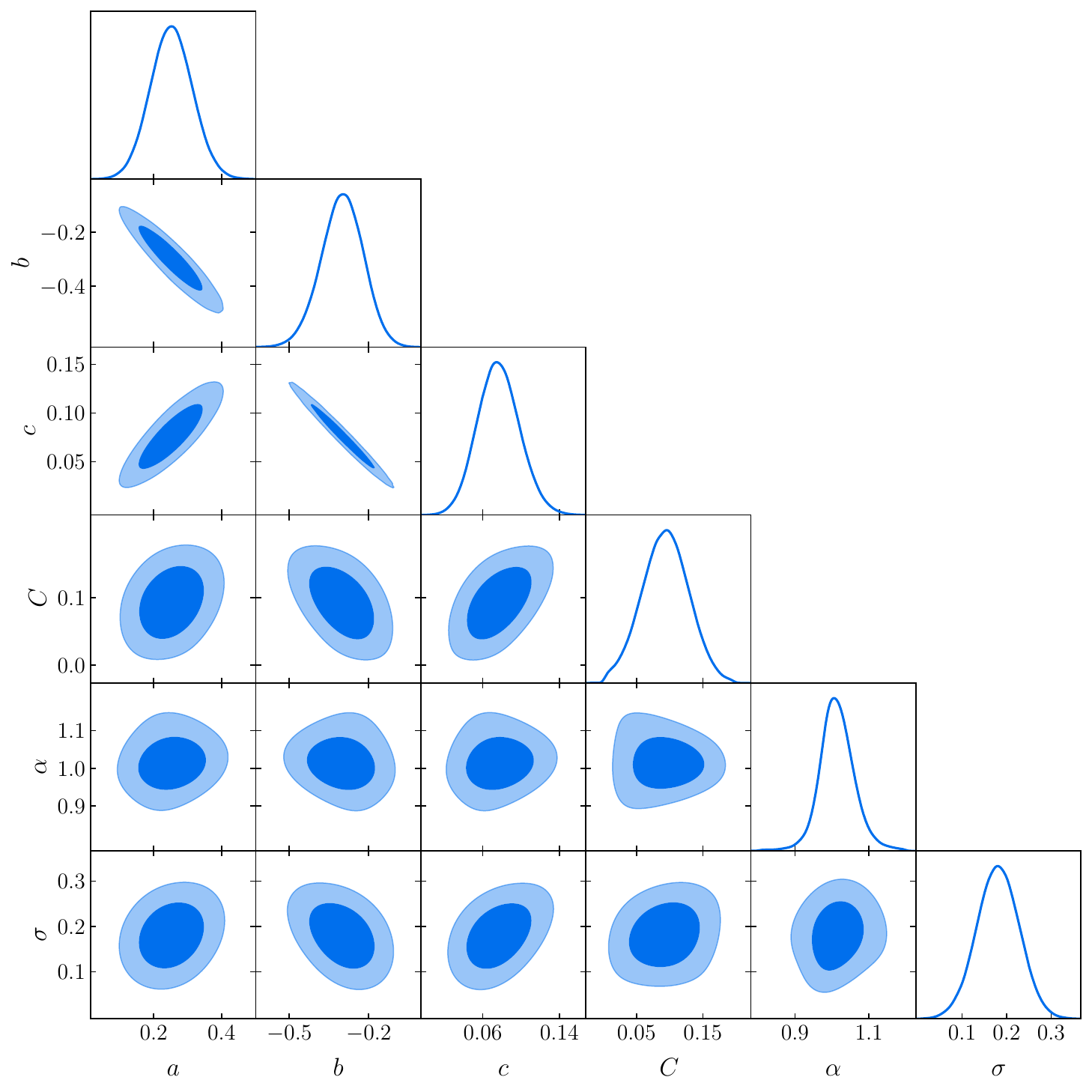}
\caption{Same as Figure~\ref{fig:contour_1715}, but for the tomographic bin at $z_{\rm eff} = 1.775$.}
\label{fig:contour_1775}
\end{figure*}

\begin{figure*}[htbp]
\centering
\includegraphics[width=0.85\textwidth]{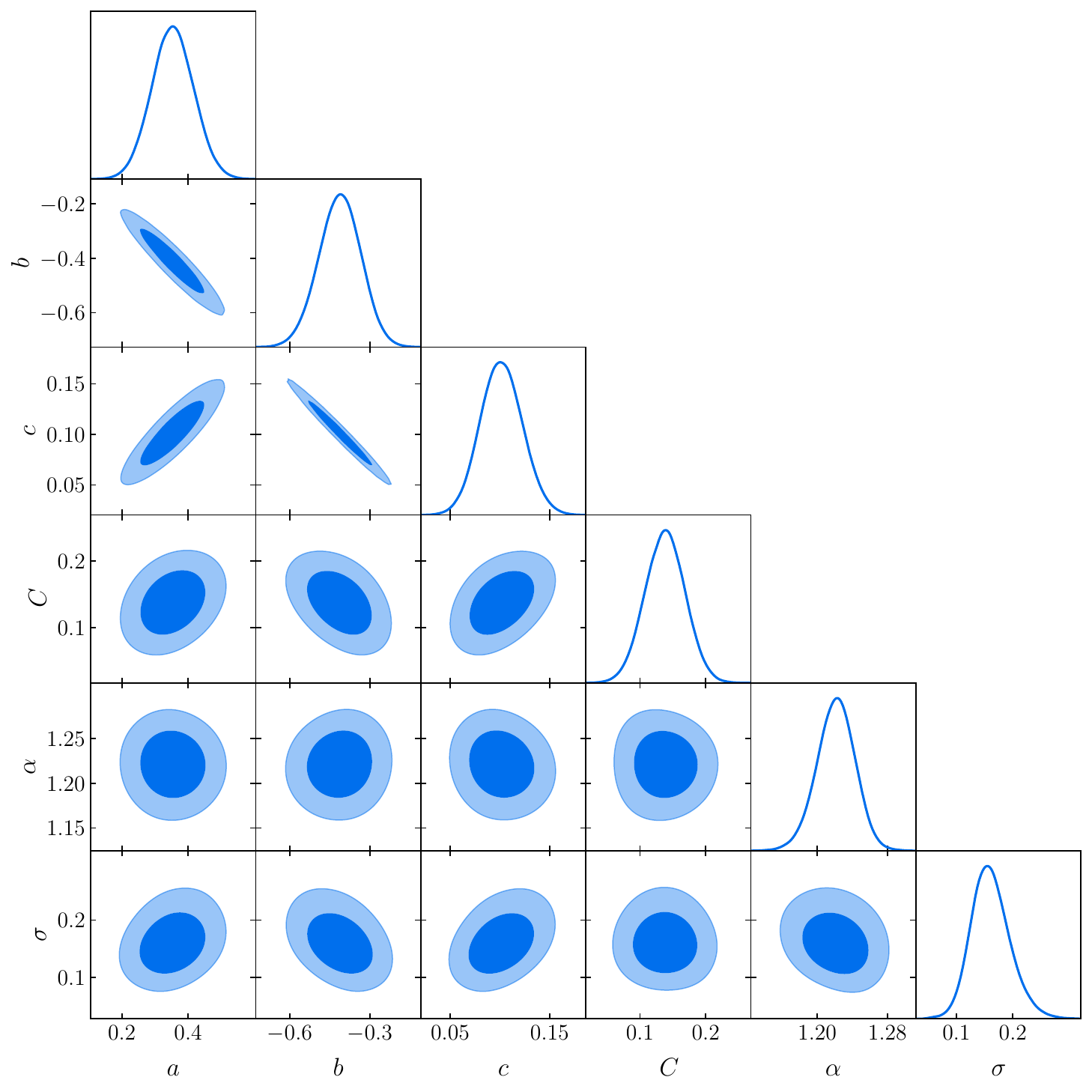}
\caption{Same as Figure~\ref{fig:contour_1715}, but for the tomographic bin at $z_{\rm eff} = 1.785$.}
\label{fig:contour_1785}
\end{figure*}

In this appendix, we present the full marginalized posterior distributions for the free parameters of our empirical transverse BAO model (equation~(\ref{eq:ajuste})). Figures~\ref{fig:contour_1715} through~\ref{fig:contour_1785} display the 1D marginalized distributions and the 2D joint confidence contours (at the $68\%$ and $95\%$ levels) obtained from the MCMC analysis for the four selected tomographic shells: $z_{\rm eff} = 1.715$, $1.725$, $1.775$, and $1.785$. 

These corner plots explicitly illustrate the intrinsic degeneracies between the broadband polynomial coefficients ($a, b, c$) and the physical BAO parameters ($C, \alpha, \sigma$). Notably, the well-constrained posteriors for the shift parameter $\alpha$ and the amplitude $C$ confirm that the acoustic peak is robustly resolved within the adopted priors for each of these high-significance bins.



\end{document}